\documentclass{nature}
\usepackage{graphicx}
\usepackage{url}
\usepackage{subfigure}
\usepackage{amsmath, amssymb, mathrsfs}
\usepackage{mathrsfs}
\usepackage{dsfont}
\usepackage{comment,color}
\usepackage{pgfplots}
\usepackage{pgf}
\usepackage{ulem}

\usepackage{lineno}

\usepackage{tikz}
\usepackage{units}
\usepackage{color}

\newcommand\apss{Astrophys. Space Sci.,\,}
\newcommand\nat{Nature,\,}
\newcommand\apj{Astrophys. J.\,}

\title{Radio sky reveals primordial electron-proton interactions }
\author{Shyam~Balaji,$^{1}$ Maura~E.~Ram\'irez-Quezada,$^{2}$ C\'eline~B\oe hm$^{3}$}

\makeatletter
\let\saved@includegraphics\includegraphics
\AtBeginDocument{\let\includegraphics\saved@includegraphics}
\renewenvironment*{figure}{\@float{figure}}{\end@float}
\makeatother

\begin{document}
\maketitle

\begin{abstract}

For several decades, astronomers have measured the electromagnetic emission in the universe from the lowest to the highest energies with incredible precision.
The lowest end of the spectrum, corresponding to radio waves, is fairly well studied and understood. Yet there is a long standing discrepancy between measurements and predictions, which has prompted the construction of many new models of radio emitters. 
Here we show that remnant electron-proton interactions, leading  to photon production in the early universe, also referred to as cosmic free-free emission, solves the discrepancy between theory and observations.  
While the possibility of cosmic free-free emission has been postulated for several decades, this is the first time that the amplitude and shape of the signal has been computed and its existence demonstrated. Using current measurements we estimate this emission to become important from around a redshift of $z \simeq 2150$. 
This contribution from fundamental particles and interactions represents the lowest energy test from the early universe of one of the pillars of modern physics, Quantum Electrodynamics. The next generation of deep radio surveys will be able to measure primordial signals from this cosmic era with greater precision and further solidify our understanding of the radio sky.
\end{abstract}

At the lowest energy tail of the electromagnetic spectrum lies radio waves. These are emitted by a number of astrophysical objects such as active galactic nuclei, radio galaxies, star forming galaxies, as well as any sources capable of producing high energy charged particles moving through diffuse galactic and inter-galactic magnetic fields\cite{Singal_2010}.  This radio background has been the subject of intense study since its discovery and imaged over large areas of the sky and at multiple frequencies\cite{deOliveira-Costa:2008cxd,1970Natur.228..847C}.
At frequencies above the radio spectrum, lies the cosmic microwave background (CMB). Since its discovery in 1965 by Wilson and Penzias, the latter has been studied in great detail. Analyses of its angular spectrum support the existence of both dark matter, dark energy (or the so-called  cosmological constant), the notion of a hot big-bang and inflation.  Early measurements of the CMB by the COBE experiment indicated that the CMB obeys a near perfect black-body spectrum; a conclusion that has remained robust for forty years\cite{Mather:1993ij}, The physics of the CMB has continued to develop, now testing both the standard model of cosmology to a higher degree of precision and constraining new physics that can manifest as deviations from current theoretical expectations. In addition, a qualitatively new cosmological probe, the physics of 21 cm HI emission/absorption at the end of the so-called Dark Ages and beginning of the comic dawn have emerged with implication in the radio domain. 

As such, it may seem that the study of cosmic electromagnetic emission in the radio-microwave range is largely completed. Yet,  deviations of the CMB spectrum have been predicted\cite{Silk:1967kq,Chluba:2011hw} and possibilities of new physics manifesting as a deviation in the Rayleigh tail\cite{Chluba:2014wda,Pospelov:2018kdh, Ghara:2021twf} have been proposed. Space-based experiments such as PIXIE\cite{Kogut:2011xw} have been planned to measure these possible deviations and will be critical to determine whether the CMB is indeed a perfect black-body spectrum.\footnote{The ARCADE experiments\cite{Fixsen:2004hp, Fixsen:2009xn} did in fact claim to find hints of deviation at a few GHz but the results have since been strongly challenged\cite{Subrahmanyan:2013eqa}.}
Furthermore, there is a noticeable discrepancy between naive extension of the CMB blackbody to low frequency and observations of radio frequencies between $10^{7}$-$10^{9}\,\rm Hz$, which deserves to be investigated and is the subject of this work.

Such a discrepancy could be due to large foreground contributions, which subsequent models have tried to address\cite{Nitu:2020vzn}
or the presence of new physics, such as dark matter decay or late-time dark matter annihilation, primordial black holes or cosmic strings\cite{Diamanti:2013bia,Pospelov:2018kdh, Ghara:2021twf}. Although they could also originate from other late time processes such as  reionisation, structure formation or magnetic fields\cite{Reis:2020arr,Vernstrom:2021hru}. Studying the $10^7-10^9$ Hz regime is thus critical to refine our understanding of modern cosmology.

The most well-known and studied model to explain this discrepancy, PB96, was developed by Protheroe and Biermann\cite{Protheroe:1996si}. In order to estimate the extragalactic radio background down to frequencies of $\rm kHz$, the PB96 model takes into account contributions from normal galaxies (NG) and radio galaxies (RG). The normal galaxies contribution is larger than that of radio galaxies in the lower frequency regime $10^{3}$-$10^{6}\,\rm Hz$ while the radio galaxies emission dominates in the $10^{6}$-$10^9\,\rm Hz$ range. However, despite its success in explaining the RG background, PB96 cannot reproduce the total radio background emission that has been observed in the $10^7$-$10^9\,\rm Hz$ regime. Other, more recent models of the CRB\cite{Nitu:2020vzn}, modulate extragalactic contributions within a larger parameter region in an attempt to match source counts.

Here we provide a more economical solution by showing that an additional radio source of primordial origin can explain the difference between the PB96 model and the observed radio background, and we quantify its magnitude for the first time. 
This contribution arises from the bremsstrahlung emission associated with electron-proton elastic scattering interactions for a very specific range of redshifts. This process is supposed to cease as the number of free electrons and protons get sufficiently suppressed. In the chronology of the cosmos, this important phase is termed ``recombination" due to the universe cooling and electrons and protons combining to form neutral Hydrogen atoms. During the recombination epoch, existing CMB photons decouple from matter at $z\simeq 1090$\cite{WMAP:2012nax}. However during this period, the free electrons and protons continue to scatter through QED interactions and emit low frequency bremsstrahlung photons. 

These photons constitute a different population from the CMB as they are not part of the thermal bath. Their energy corresponds to the radio range as they get redshifted and diluted by cosmic expansion. 
This same band of photons can also be efficiently absorbed by the same interaction in reverse. Here we show however that for a very specific range of redshifts, the bremsstrahlung photons emitted by this free-free emission are not completely absorbed by the reverse process. As a result, many of these photons from primordial origins contribute to the radio background today. More surprisingly even, this small contribution can fit the radio discrepancy between theory and diffuse radio flux measurements precisely\cite{Gervasi:2008rr}.

\section*{ Calculating the number of photons emitted and absorbed by electron-proton interactions } 

Recombination, is one of the most important phases in the history of the universe. 
Since it is not instantaneous, free protons and electrons can continue to interact and produce low energy photons through the radiative process  $e + p \to e + p +\gamma$.
As the photon energy reduces, the probability of emission increases. In light of this, one expects this process to be theoretically infrared (IR) divergent, predicting production of an infinite number of photons. A result that is clearly not physical. We will see in subsequent sections that within the context of Quantum Field Theory (QFT) such an IR divergence is cancelled by the reverse process, ensuring that only a finite number of photons can be observed. Hence the study of such a channel provides, in principle, a remarkable window into QED at low energy and where the IR cancellation takes place.
Furthermore, we will see that the bremsstrahlung photons considered are spectrally separated from the usual CMB photons. Therefore, they do not thermalise into the same blackbody spectrum and can persist as a distinct non-thermal population in the present universe's radio sky.

To determine the fate of these emitted photons, one needs to solve a set of coupled Boltzmann equations to follow the evolution of the energy distributions of electron, proton and photon wherein both the $e + p \to e + p +\gamma$ and $e + p +\gamma \to e + p $ processes should be included. However, since the focus of this letter is on photons and ordinary matter is meant to be fully thermalised during recombination, it suffices to compute the evolution of the photon distribution. For details, please refer to Sec.~\ref{app:Emiss_rate_details} in Methods. 

The subtlety in such a calculation, is to accurately describe the population of photons that could be reabsorbed by the reverse process. To do this, we first compute the differential number of photons of energy $E_\gamma$ produced by free-free emission $\frac{\partial^2n_{\rm ems}}{\partial z \partial E_\gamma}$, and remove the photons through the absorption term $\frac{\partial^2n_{\rm abs}}{\partial z \partial E_\gamma}$. Considering that electrons and protons follow a Maxwell-Boltzmann distribution, these two quantities are given by
\begin{equation} 
\frac{\partial^2 n_\textrm{ems}}{\partial z \partial E_\gamma}=\frac{\partial Q_{\rm ems}}{\partial 
E_\gamma}\mathcal{J}(z),\ \,\,\,\,     
\frac{\partial^2\tilde{n}_{\rm abs}}{\partial z \partial 
E_\gamma}=  \frac{\partial Q_{\rm abs}}{\partial E_\gamma}\mathcal{J}(z),\ \,\,\,\,  
\label{eq:n_photon_emss_abs}
\end{equation}
where $n_\gamma$ refers to the net photon density and the Jacobian factor $\mathcal{J}(z)=\frac{1+z}{T}\left|\frac{dt}{dz}\right|$. The rates $\frac{\partial Q_{\rm ems}}{\partial E_\gamma}$ and $\frac{\partial Q_{\rm abs}}{\partial E_\gamma}$ are the results of phase space integrals which are given in Eqs.~\eqref{eq:emission_rate_final} and \eqref{eq:absorption_rate_final} respectively.
We define 
\begin{equation} \frac{\partial^2n_\textrm{abs}}{\partial z \partial E_\gamma}=f_\gamma(E_\gamma)\,\frac{\partial^2\tilde{n}_\textrm{abs}}{\partial z dE_\gamma}, \ \  {\rm{with}} \ \  f_\gamma(E_\gamma)=\frac{\pi^2}{\left((1+z)E_\gamma\right)^2} \frac{\partial n_{\rm \gamma}}{\partial  E_\gamma},
\end{equation} 
  where $f_\gamma$ is the non-thermal photon distribution function related to the net photon emission number density. 
Since we are interested in the evolution of the number density of photons emitted over time, we  have expressed the photon number density  in terms of the redshift $z$.  This is achieved by using the relationship between the proper time $t$ and$z$\cite{Condon:2018eqx,Macdonald:2006sa} with matter and dark energy densities of $\Omega_M=0.265$\cite{Scott:2004pr,Planck:2018vyg}and $\Omega_\Lambda=0.685$\cite{Planck:2018vyg} respectively.

Eventually the number of photons that are expected to contribute to the radio background is  given by the balance of these two terms. The integro-partial differential equation describing the net observable photon number density is given by
\begin{align}
 \frac{\partial^2n_{\rm \gamma}}{\partial z \partial E_\gamma}=\left[\frac{\partial^2 n_\textrm{ems}}{\partial z \partial E_\gamma}-f_\gamma(E_\gamma)\frac{\partial^2\tilde{n}_\textrm{abs}}{\partial z \partial E_\gamma}\right] a(z)^3\label{eq:n_net},
\end{align}
with boundary condition $\frac{\partial n}{\partial E_\gamma}(z_\textrm{neq})=0$, where $z_\textrm{neq}$ is the redshift at which the free-free photons begin to accumulate as a distinct non-thermal population. The first term on the right handside corresponds to photon injection while the second corresponds to absorption. The necessary physical condition that absorption can never exceed emission is ensured dynamically by $f_\gamma$. Finally, we have taken into account the expansion of the universe by introducing the cosmic scale factor $a(z)$ for dilution. We then solve Eq.~\eqref{eq:n_net} to obtain $\frac{\partial n}{\partial E_\gamma}$ as a function of redshift $z$.

\section*{Predicting the radio background associated with $e p \rightarrow e p \gamma$ }
\label{sec:results}

\begin{figure}[!ht]
    \centering
      \includegraphics[width=0.47\textwidth]{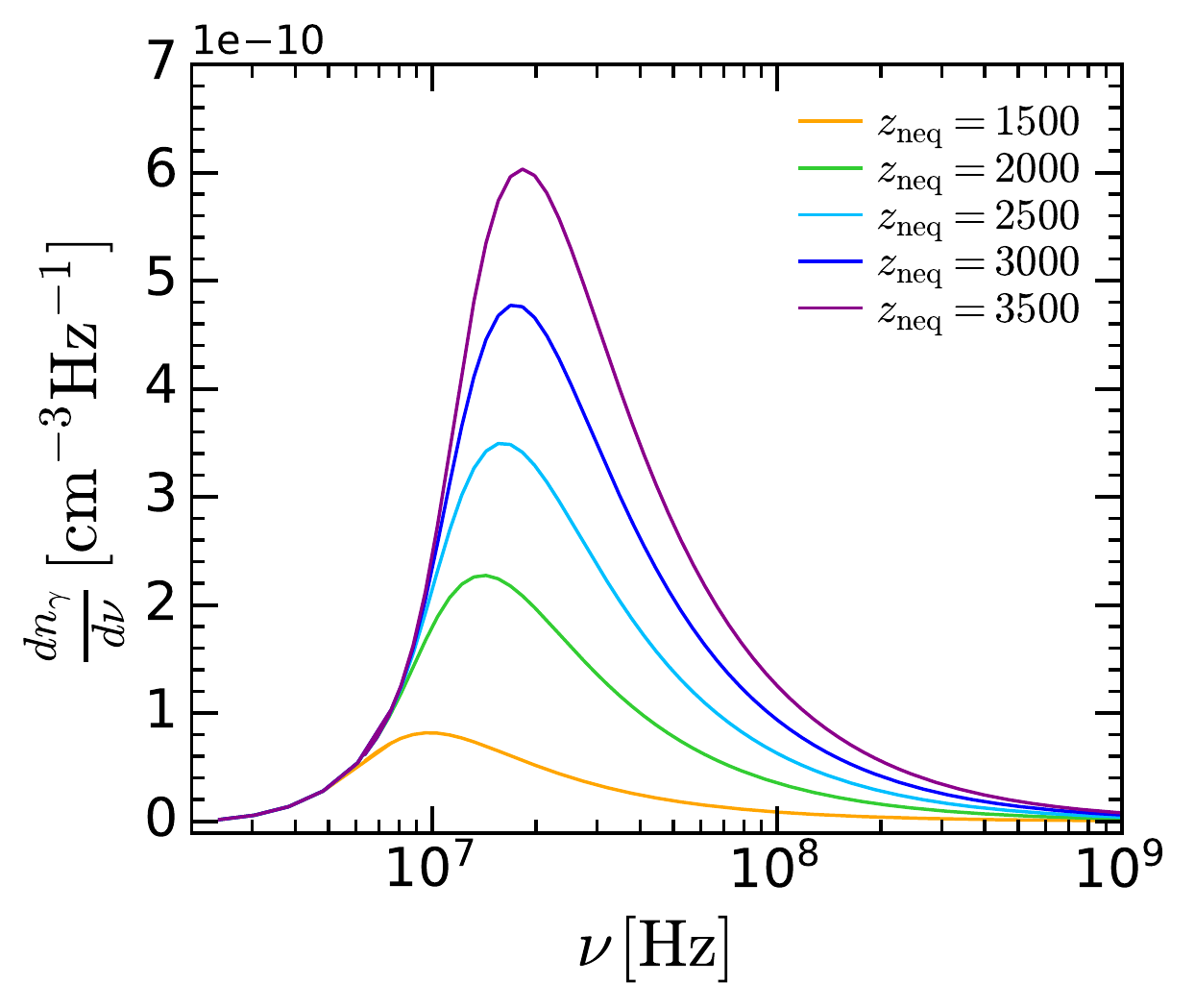}
         \includegraphics[width=0.5\textwidth]{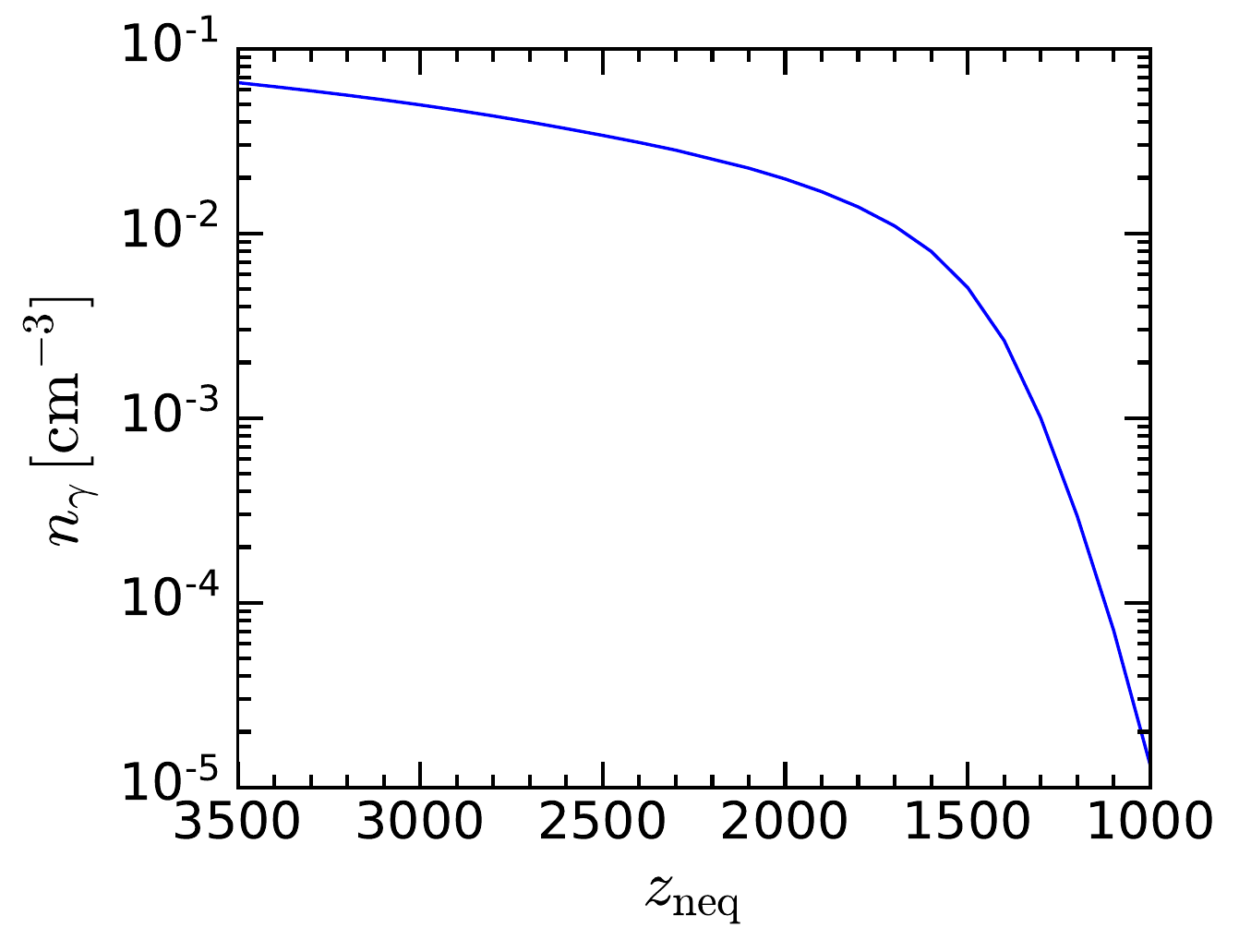}
     \caption{Differential photon number density  as a function of frequency at present time $z=0$ (left panel). Each colour corresponds to different decoupling times varying from $z_{\rm neq}=3500$ (purple line) to $z_{\rm neq} = 1500$ (orange line). The free-free photon number density at present time, is shown as a function of the decoupling redshift $z_{\rm neq}$ (right panel).}
    \label{fig:dndw_ODE_log}
\end{figure}

The number density of photons expected today as a result of the emission and absorption process after recombination is plotted in Fig.~\ref{fig:dndw_ODE_log} as a function of frequency (left panel) and out-of-equilibrium redshift $z_{\rm neq}$ (right panel).  The fact that $n_{\gamma}$ is significant (at most $\mathcal{O}(10^{-1})\, \rm cm^{-3}$) indicates that the emission rate dominates over the absorption rate at radio frequencies which manifests in the radio background estimated in this letter. The differential number density peaks somewhere between $10^7$-$10^8\,\rm Hz$  but more importantly, we note that it  becomes negligible for frequencies below $10^7 \, \rm{Hz}$. This is a very important result. In fact, theoretically, the number of photons emitted at extreme low frequencies is expected to diverge to infinity, but this occurs for absorption as well. Using QFT, we are able to demonstrate that both contributions actually cancel out each other at frequencies smaller than $10^7 \, \rm{Hz}$ due to $f_\gamma$. Furthermore for $z_{\rm neq} \simeq 1000$, the expected number density falls down to $\mathcal{O}(10^{-5}) \,\rm cm^{-3}$. This is unsurprising given the rapid reduction in electrons and protons forming neutral hydrogen during recombination. Hence, we observe three distinct regimes. At frequencies greater than $10^{10} \, \rm{Hz}$, the number of bremsstrahlung photons emitted by free-free interactions is negligible. In the $10^7$-$10^{10}\,\rm Hz$ range, both the emission and absorption increase but the emission is greater than the absorption. Finally for frequencies smaller than $10^7 \, \rm{Hz}$, both the emission and absorption terms diverge but they also cancel each other out. These regimes in conjunction leads to a finite result in a definite radio frequency range. Measurement of this signal should enable precise determination of when the free-free photons in the universe left thermal equilibrium, and ultimately should allow robust constraints to be placed on $z_{\rm neq}$.

We note that classical estimates of a possible radio contribution were mentioned in early studies of the CMB\cite{1975ApJ...195....1C,1969Ap&SS...4..301Z}. However this contribution was deemed negligible at frequencies greater than a few $\rm GHz$, where the CMB signal is dominant\cite{PhysRevD.48.485}. Furthermore, at the time, radio sources in the foreground were modelled poorly, which explains why this contribution has been forgotten until now. Finally, it is worth mentioning that classical free-free interaction estimates are plagued by divergent IR behaviour which is in turn also regulated with arbitrary cutoff scales\cite{Brussaard1962,Chluba:2019ser}. As such the classical approach is sufficient for crude estimates of the magnitude of the effect, but it doesn't provide a precise physical spectrum, particularly at low frequency. Using a full QFT treatment has enabled us to solve this issue and, as a result, we provide the first estimate of the radio background emitted by bremsstrahlung emission associated with cosmic free-free interactions.\footnote{We have disregarded other process such as $e$-$e$ and $p$-$p$  bremmstrahlung as they are greatly suppressed due to having the same mass.}

\section*{Relevance of this novel radio background for current observations}

\begin{figure}[!ht]
\centering
  \includegraphics[width=0.5\textwidth]{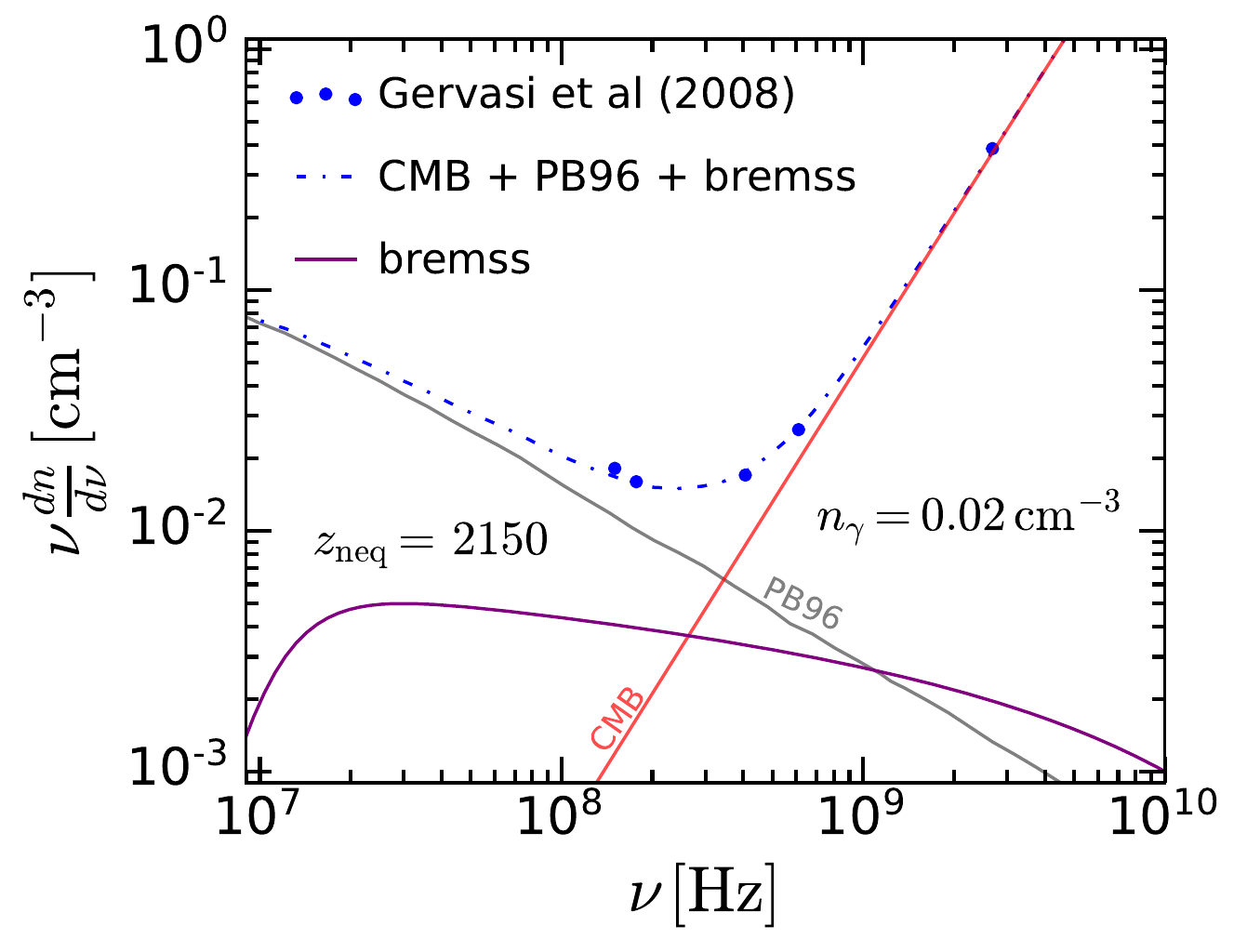}
    \includegraphics[width=0.49\textwidth]{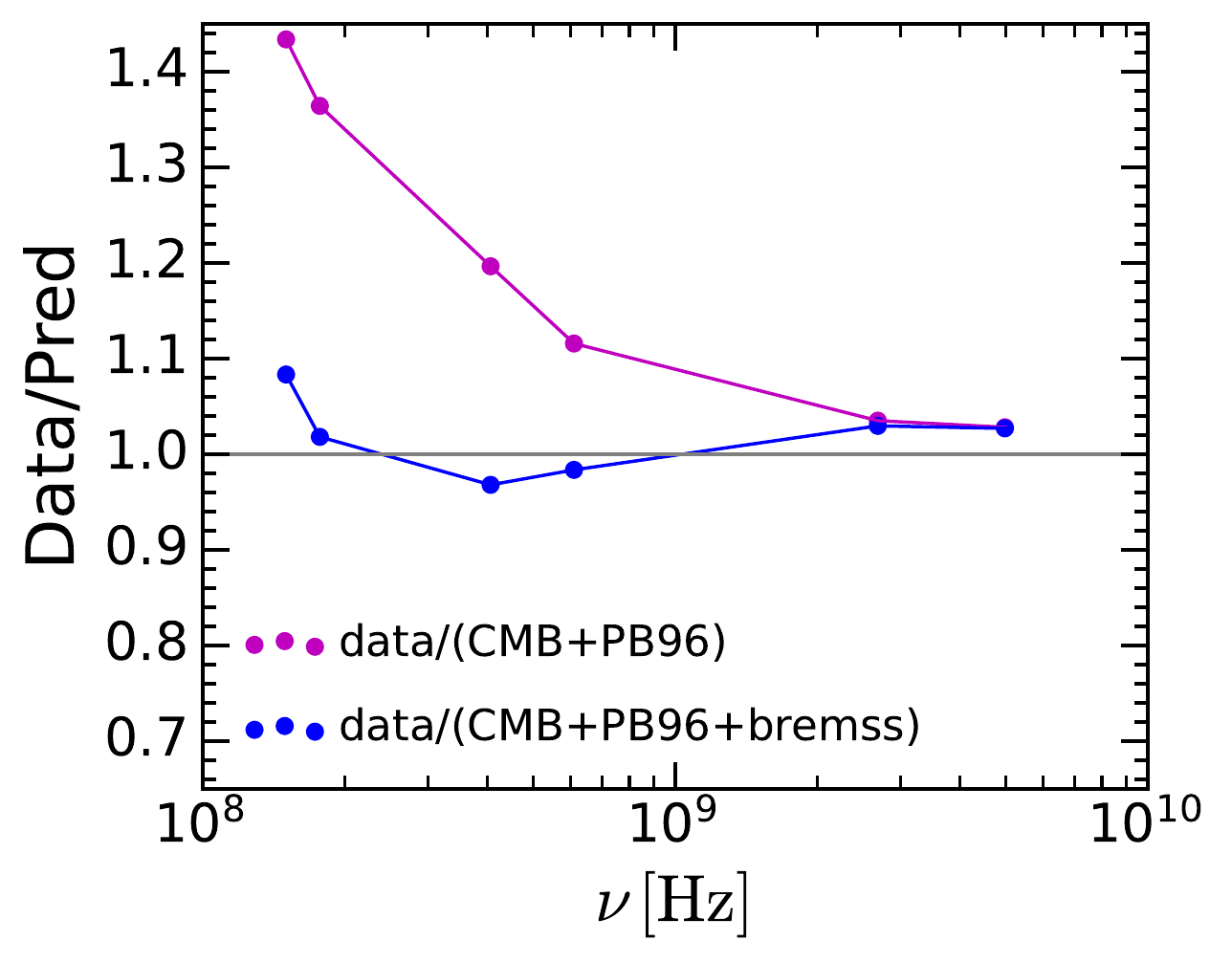}
         \caption{Left panel: Differential photon number density by frequency vs frequency. We show the  total theoretical prediction (blue dashed line) and data\cite{Gervasi:2008rr} (blue points). We also show PB96 foreground model (gray line), the CMB blackbody (red line) and recombination bremsstrahlung decoupling at $z_{\rm neq}=2150$ (purple line). Right panel: The ratio between data\cite{Gervasi:2008rr} and the prediction with (blue) and without (magenta) the recombination bremsstrahlung contribution.}
    \label{fig:dndw_CMB_bremss}
\end{figure}

To understand the relevance of this background, we introduce foreground models and radio observations. We shall use the PB96 model\cite{Protheroe:1996si}, as it is the most well-known and studied that intends to explain the CRB foreground. In order to estimate the radio foreground down to frequencies of $\rm kHz$ the PB96 model takes into account contributions from normal and radio galaxies.  The model also includes free-free emission and  absorption by the warm and hot ionized components of the interstellar medium and by synchrotron self-absorption.

In the left panel of Fig.~\ref{fig:dndw_CMB_bremss}, we show that the cosmic free-free contribution considered in this work can resolve the tension observed between PB96 and radio source count\cite{Gervasi:2008rr}. The CMB black-body is shown in red and is computed using Eq.~(14) of~\cite{Hill:2018trh} while PB96 predictions of galactic foregrounds are shown in gray. The free-free contribution discussed in this work is shown in purple and the sum of all relevant contributions is shown in blue dot-dashed. Inspecting the right panel of Fig.~\ref{fig:dndw_CMB_bremss}, we see the ratio between model predictions and data can exceed 43\%. We find that if free-free photons start manifesting as a separate population from CMB photons from $z_{\rm neq}\simeq2150$, we can reduce the tension with data to at most 8\%. The number density in the present universe corresponding to this is $n_\gamma=0.024\,\textrm{cm}^{-3}$, which is very small relative to the CMB number density of approximately $411\,\textrm{cm}^{-3}$, but still represents around 5\% of the radio foreground contribution considered. The result also provides a definite prediction for the expected radio spectrum between the frequencies of the observed source counts in the range $10^7<\nu < 10^9\,\rm Hz$ which will be tested at high precision by the Square Kilometre Array and the Expanded Very Long Array in the near future. As explained above, frequencies below $10^7\,\rm Hz$  receive  negligible correction from free-free contributions and therefore no significant excess radio signal from this process is expected in this region.

\section*{\label{sec:conclusion}Conclusion}
We have shown that the bremsstrahlung emission associated with free-free electron-proton scattering in the early universe constituted a significant source of radio emission today and explains the discrepancy between models of radio sources and observations. 

Numerical estimates of the photon net number density indicate that if free-free photons decoupled from the thermal bath at $z_{\rm neq}\approx2150$, the tension between observational data in \cite{Gervasi:2008rr} and the theoretical predictions can be significantly reduced, from 43\%, 36\%, 19\% and 11\% to 8\%, 1\%, 4\% and 2\% at frequencies of $151\,\rm MHz$, $178\,\rm MHz$, $408\,\rm MHz$ and $610\,\rm MHz$  respectively. The photon number density at present time for this signal is $ n_\gamma \simeq 0.02\,\rm cm^{-3}$, which is very suppressed compared to that of the CMB but not insignificant relative to the radio sky. Finally we see that frequencies below $10^7\,\rm Hz$ and above $10^{10}\,\rm Hz$ receive negligible corrections from recombination bremsstrahlung and hence, no significant excess radio signal from this process is expected in these regions. Therefore, our work provides a robust test of QED from recombination in the universe down to temperatures of about 0.1 eV. The Square Kilometre Array and Expanded Very Long Array are expected to probe intermediate frequencies with high resolution which would further strengthen our understanding of this process. Finally, discovery of potential nano-Jansky populations and radio relics from clusters as additional sources of faint diffuse emission will enable more precise data driven extraction of $z_{\rm neq}$ in the future.  


\begin{addendum}
\item[Correspondence]Correspondence and requests for materials should be addressed to S. Balaji (email: sbalaji@lpthe.jussieu.fr).
\item[Statement of contribution] The idea was proposed by C. B\oe hm. S. Balaji derived the solution for the dynamic number density. M.E. Ramirez-Quezada produced all the displayed plots. S. Balaji and M.E. Ramirez-Quezada identified the dominant channel and performed all calculations and analysis. All authors have contributed to all stages of the project and the writing of the manuscript.

\item[Acknowledgements]We would like to thank Anastasia Fialkov and Matthew Johnson for helpful discussions at the beginning of the project. We would also like to thank Ravi Subrahmanyan for useful discussions regarding radio source counts. CB and MRQ would like to thank the Perimeter Institute for their hospitality when the project started and CB would like to thank Laurent Freidel for stimulating discussions on QFT at low energy. SB is supported by 
funding from the European Union's Horizon 2020 
research and innovation programme under grant 
agreement No 101002846 (ERC CoG ``CosmoChart'') 
as well as support from the Initiative Physique 
des Infinis (IPI), a research training program of the Idex SUPER at Sorbonne
Universit\'{e}. MRQ is supported  by JSPS KAKENHI Grant Number 20H01897. 
\item[Author Information] \begin{affiliations}
\item Laboratoire de Physique Th\'{e}orique et Hautes Energies (LPTHE), UMR 7589 CNRS \& Sorbonne Universit\'{e}, 4 Place Jussieu, F-75252, Paris, France
\item Department of Physics, University of Tokyo, Bunkyo-ku, Tokyo
 113--0033, Japan
 \item School of Physics, The  University of Sydney, NSW 2006, Australia
\end{affiliations}

\item[Conflicts of interest] The authors declare that they have no competing financial interests.

\end{addendum}
\newpage
\listoffigures

\begin{methods}
\section{\label{app:Emiss_rate_details} Emission rate computation}

We  compute the  rate of bremsstrahlung emission in electron-proton scattering  $e(p_1) + p(p_2)\to e(p_3) + p(p_4)+\gamma(k)$.
The photon emission rate $Q_\textrm{ems}$ from this process is generically given by\cite{Dent:2012mx},
\begin{align}
     Q_\textrm{ems} &= d\Pi_\gamma \int  d\Pi_1 d\Pi_2 d\Pi_3 d\Pi_4 \mathcal{S} |\mathcal{M}_\gamma|^2\delta^4(p_1+p_2-p_3-p_4-k)  f_1 f_2 (1-f_3), (1-f_4)\label{eq:emission_rate_definition}
\end{align}
where, $|\mathcal{M}_\gamma|^2$ is the squared amplitude for bremsstrahlung emission, the symmetry factor $\mathcal{S}=1$ for $e$-$p$ scattering, $d\Pi_j=d^3\mathbf{p}_j/\left(2E_j(2\pi)^3\right)$ are the particle phase space factors and $f_j$ denotes the distribution function of the particles involved in the process.  In the low energy limit and ignoring Pauli blocking, we take $(1-f_{3,4})\to 1$ as well as $E_{1,3}=m_e$ and $E_{2,4}=m_p$. Since the photon abundance is small compared to
the particles in the thermal plasma, we can neglect $f_\gamma$ in $Q_{\rm ems}$.
 For non-relativistic particles, the distribution function for electrons and protons is Maxwell-Boltzmann
\begin{align}
f(p_j) &= \frac{n_j}{2}\left(\frac{2\pi}{m_j T}\right)^{3/2} e^{-\mathbf{p}_j^2/2m_j T},
\end{align}
where $n_j$ and $m_j$ are the number density and mass of the particles in the process. Finally, the squared amplitude of free-free emission ($ep \rightarrow ep\gamma$) is given by,
\begin{equation}
   \sum_{\rm spins}|\mathcal{M}_\gamma|^2=e^2\sum_{\rm spin}|\mathcal{M}_0|^2\left[\frac{p_3}{p_3\cdot k}-\frac{p_1}{p_1\cdot k}\right]^2,
\end{equation}
where $\mathcal{M}_0$ corresponds to $e$-$p$ scattering. The total expression in terms of variables $u$ and $v$, defined in Eq.~\eqref{eq:u_v_deffinitions}, is therefore given by,
\begin{align}
    \sum_{\rm spins}|\mathcal{M}_\gamma|^2
    &=\frac{2 e^6}{m_e^2\mu T^3x^2
    (u+v-2z\sqrt{uv})}\bigg[\!\!\left(m_e^2+m_p^2+2 \mu  T (u+v)\right)^2-4 \mu ^2 T^2 (u-v)^2\nonumber\\
    &+\!\left(m_e^2+m_p^2+\mu  T \left(2 z \sqrt{u v}+u+v\right)\right)^2\!-2 \mu  T \left(m_e^2+m_p^2\right) \left(-2 z \sqrt{u v}+u+v\right)\!\!\bigg]
\end{align}

In order to compute the emission rate we make the useful change of variables $l=p_2-p_3$, $r=p_2-p_4$\cite{Dent:2012mx}. Taking the photon soft limit, for these parameters leads to
\begin{align}
    \mathbf{r}^2 &=\mu T (u+v-2y\sqrt{uv}),\nonumber\\
    \mathbf{l}^2 &=\mu T (u+v+2y\sqrt{uv}),\nonumber\\
    \mathbf{r}\cdot \mathbf{l} &= \mu T(u-v),
\end{align}
where the reduced mass is given by $\mu\equiv \frac{2m_pm_e}{m_e+m_p}$.  The new variables  $y$, $u$ and $v$ are dimensionless and are defined in terms of the temperature by
\begin{equation}
     y=\cos\theta,\,\,u=\frac{\mathbf{p}_i^2}{\mu T},\,\,v=\frac{\mathbf{p}_f^2}{\mu T}.\label{eq:u_v_deffinitions}
\end{equation} 
which corresponds to the angle $\theta$ between the initial ($\mathbf{p}_i$) and final ($\mathbf{p}_f$) 3-momentum of the particles in the centre of momentum coordinates,
\begin{equation}
    \mathbf{p}_{1,2}=\mathbf{P}\pm \mathbf{p}_i,\,\,\mathbf{p}_{3,4}=\mathbf{P}'\pm \mathbf{p}_f.
    \end{equation}

 From here, we rewrite the product of the distribution functions $f_1f_2$ in Eq.~\eqref{eq:emission_rate_definition} as
\begin{equation}
    f_1f_2\to \mathrm{exp}\left[-\frac{\mathbf{P}^2}{\mu T}-\frac{\mathbf{p}_i^2}{\mu T}-\left(\frac{1}{m_e T}-\frac{1}{m_pT}\right)\mathbf{P}\cdot \mathbf{p}_i\right]\label{eq:new_f}.
\end{equation}
We may then rewrite the integration variables in Eq.~\eqref{eq:emission_rate_definition} in terms of 3-momentum integrals over $\mathbf{p}_i, \mathbf{P}$ and $\mathbf{p}_f$ which can then be written as integrals in terms of the dimensionless $u$, $v$ and $y$ like
\begin{align}
   & \int d^3\mathbf{p}_f= \pi(\mu T)^{3/2}\int\sqrt{v}dv\int_{-1}^1dy,\\
   &\int d^3\mathbf{P} f_1 f_2=\frac{\pi (\mu T)^{3/2}}{m_p^2} f(u),\\
   &\int d^3\mathbf{p}_i = 4\pi\int p_i^2 d\mathbf{p}_i = 2\pi(\mu T)^{3/2} \int \sqrt{u} du
\end{align}
where $f(u)$ is obtained after integration by replacing the definitions in Eq.~\eqref{eq:u_v_deffinitions} into Eq.~\eqref{eq:new_f}, the explicit expression is given by
 \begin{align}
     f(u)\!=\!\!\bigg[2e^{-u}m_p\sqrt{u}(\mu-m_p)+e^{\frac{u\mu(\mu-2m_p)}{m_p^2}}\!\!\sqrt{\pi}(m_p^2(1+2u)-u(4m_p-2\mu)\mu)\mathrm{erfc}\!\!\left(\frac{\sqrt{u}(\mu-m_p)}{m_p}\right)\!\!\bigg]
     \label{eq:functionu}
\end{align}
where $\textrm{erfc}$ is the complimentary error function.
Finally the differential photon emission rate is 
\begin{align}
     \frac{\partial Q_\textrm{ems}}{\partial x}&=\frac{\alpha_{em}^3 \mu ^{7/2} n_p n_e }{4\pi  (m_e m_p)^{11/2} \sqrt{T} x} \int dudvdy\delta(u-v-x)\sqrt{uv}f(u)\mathcal{I}(u,v,y)\label{eq:emission_rate_final}
\end{align}
where $n_e$ and $n_p$ are the electron and proton number densities respectively and $x=E_\gamma(1+z)/T$, where we define $E_\gamma$ at redshift $z=0$ and $x$ at arbitrary $z$.  Notice that we have integrated over $\textbf{p}_4$ already. The function $\mathcal{I}(u,v,y)$ is related to the squared amplitude for bremsstrahlung in terms of variables $u$ and $v$,
\begin{align}
    \mathcal{I}(u,v,y)\!=\!
  \frac{1}{(u\!+\!v-2y\sqrt{uv})}&\!\bigg[\!\!\left(m_e^2+m_p^2+2 \mu  T (u+v)\right)^2\!+\!\left(m_e^2+m_p^2+\mu  T \left(2 y \sqrt{u v}\!+\!u\!+\!v\right)\right)^2\nonumber\\
  &-2 \mu T \left(m_e^2+m_p^2\right) \left(-2 y \sqrt{u v}+u+v\right)-4 \mu ^2 T^2 (u-v)^2\bigg] \label{eq:I_expression}.
\end{align}
We use  the delta function to integrate over $v$ leading to the replacement $v\to u-x$.
\section{\label{app:Abs_rate_details} Absorption rate computation}
We  compute  photon absorption rate in electron-proton scattering  $e(p_1) + p(p_2)+\gamma(k)\to e(p_3) + p(p_4)$. 
The differential photon absorption rate is calculated in a very similar fashion to the emission rate. 
Notice that for the case of photon absorption ($ep\gamma\rightarrow ep$), we can directly compute the squared scattering amplitude for the inverse process by replacing $k\to-k$
\begin{equation}
    \sum_{\rm spins}|\mathcal{M}_\gamma(k)|^2=\sum_{\rm spins}|\mathcal{M}'_\gamma(-k)|^2
\end{equation}
leading to
\begin{equation}
   \sum_{\rm spins}|\mathcal{M}'_\gamma|^2=e^2\sum_{\rm spin}|\mathcal{M}_0|^2\left[\frac{p_1}{p_1\cdot k}-\frac{p_3}{p_3\cdot k}\right]^2.
\end{equation}
where $\mathcal{M}'_\gamma$ is the amplitude for absorption, which takes a similar form as the amplitude for the emission process.

Recasting with the same dimensionless variables as before, we define the differential photon absorption rate by\cite{Dent:2012mx}
\begin{align}
     \frac{\partial Q_\textrm{abs}}{\partial x}&=\frac{\alpha_{em}^3 \mu ^{7/2} n_p n_e }{4\pi  (m_e m_p)^{11/2} \sqrt{T} x}\int dudvdy\delta(u-v+x)\sqrt{uv}\mathcal{I}(u,v,y)f(u)\label{eq:absorption_rate_final}.
\end{align}
 Finally, the expression for the $\mathcal{I}$ function in the emission and absorption rates is given by Eq.~\eqref{eq:I_expression}. The delta function in the absorption rate leads to the replacement of $u\to v-x$.
\end{methods}
\end{document}